\documentclass{ifacconf}

\usepackage{graphicx}      
\usepackage{natbib}        
\usepackage{amsmath}
\usepackage{multirow}
\usepackage{textcomp}
\makeatletter
\def\endfigure{\end@float}
\def\endtable{\end@float}
\makeatother

\let\ifacconfcaptionwidth\captionwidth
\usepackage[caption=false]{subfig}
\let\captionwidth\ifacconfcaptionwidth

\begin{document}
\begin{frontmatter}

\title{Deep Learning-Based Vehicle Speed Prediction for Ecological Adaptive Cruise Control in Urban and Highway Scenarios\thanksref{footnoteinfo}}  

\thanks[footnoteinfo]{This work is funded by the German Ministry for Education and Research (BMBF) and partially supported by the Center of Commercial Vehicle Technology (Zentrum für Nutzfahrzeugtechnologie, ZNT) at the University of Kaiserslautern.}

\textbf{Sai Krishna Chada* ~Daniel G{\"o}rges* ~Achim Ebert**  ~Roman Teutsch***}

\textit{~*~Institute of Electromobility ~**~Human Computer Interaction Group ~***~Institute for Mechanical and Automotive Design \\ 
University of Kaiserslautern, Germany} \\
\textit{chada@eit.uni-kl.de, goerges@eit.uni-kl.de, ebert@cs.uni-kl.de \& teutsch@mv.uni-kl.de}

\begin{abstract}                
In a typical car-following scenario, target vehicle speed fluctuations act as an external disturbance to the host vehicle and in turn affect its energy consumption. To control a host vehicle in an energy-efficient manner using model predictive control (MPC), and moreover, enhance the performance of an ecological adaptive cruise control (EACC) strategy, forecasting the future velocities of a target vehicle is essential. For this purpose, a deep recurrent neural network-based vehicle speed prediction using long-short term memory (LSTM) and gated recurrent units (GRU) is studied in this work. Besides these, the physics-based constant velocity (CV) and constant acceleration (CA) models are discussed. The sequential time series data for training (e.g. speed trajectories of the target and its preceding vehicles obtained through vehicle-to-vehicle (V2V) communication, road speed limits, traffic light current and future phases collected using vehicle-to-infrastructure (V2I) communication) is gathered from both urban and highway networks created in the microscopic traffic simulator SUMO. The proposed speed prediction models are evaluated for long-term predictions (up to $10\,\mathrm{s}$) of target vehicle future velocities. Moreover, the results revealed that the LSTM-based speed predictor outperformed other models in terms of achieving better prediction accuracy on unseen test datasets, and thereby showcasing better generalization ability. Furthermore, the performance of EACC-equipped host car on the predicted velocities is evaluated, and its energy-saving benefits for different prediction horizons are presented. 
\end{abstract}


\end{frontmatter}

\section{Introduction}
Vehicle speed prediction is regarded as a key aspect in intelligent transportation systems (ITS) that fundamentally aim at improving the road safety, traffic-efficiency and vehicular energy efficiency \citep{Jiang2017}. Past studies on speed forecasting in transportation systems focused mainly in two directions, namely, network-wide traffic speed prediction \citep{Cui2018}, and host vehicle velocity prediction \citep{Sun2015,Gaikwad2019}. Forecasting the future velocities of the host vehicle for the entire driving route for the purpose of efficient energy management in hybrid electric vehicles (HEVs) is studied in \citep{Sun2015}. Moreover, the vehicle speed prediction can largely benefit the newly developed advanced driver assistance systems (ADAS) as well \citep{Schmied2015}. With a goal to minimize the energy consumption in on-road vehicles, efforts are being made towards developing advanced adaptive cruise control (ACC) concepts to control the host vehicle in an automated fashion. In this regard, ecological adaptive cruise control (EACC) concepts, that use an optimal controller to compute an energy-optimal speed for the host vehicle while tracking a target (leading) vehicle are becoming popular \citep{Moser2015}. In prior studies  \citep{Weissmann2018,Chada2020}, to explore the energy consumption reduction benefits using EACC in a car-following scenario, a common assumption was made that the future velocities of the target vehicle are perfectly available. However, the perfect future velocities of the target vehicle in a real world setting are not known a priori, but rather must be predicted through either behavioral models or data-driven approaches. 

Several methods for developing vehicle speed predictors for various applications were proposed in the prior works. For instance, a comparative analysis on the parametric and non-parametric approaches for speed prediction in highway driving is presented in \citep{Lefevre2014}. The authors classified the prediction space into short-term prediction ($<4\,\mathrm{s}$) and long-term prediction ($4-10\,\mathrm{s}$). Vehicle velocity prediction until $10\,\mathrm{s}$ for the use case in energy management in HEVs is studied in \citep{Gaikwad2019}. \citep{Liu2019} investigated on the one hand stochastic models such as Markov chain and conditional linear Gaussian (CLG) for the host vehicle velocity prediction. On the other hand, deterministic models such as auto-regressive moving average (ARMA), nonlinear auto-regressive exogenous model (NARX) and recurrent neural networks such as long-short term memory (LSTM) units were studied. The authors in \citep{Wegener2021} studied longitudinal vehicle speed prediction in urban environments using CLG and deep neural networks (DNNs). Moreover, \citep{Shin2019} proposed a fuzzy markov chain model with speed constraints to perform host vehicle speed prediction. In \citep{Lin2018}, the authors proposed a cloud-based seasonal autoregressive integrated moving average (SARIMA) framework for vehicle speed prediction purpose, for which a highway database was used. To enhance the accuracy of the speed predictors, availing the benefits of the surrounding information using vehicle-to-vehicle (V2V) and vehicle-to-infrastructure (V2I) communication can be essential \citep{Moser2015}. 


Concerning the speed prediction for the ecological cruise control use case, there exist only a handful of studies \citep{Schmied2015,Moser2015,Jia2020,Sankar2022}. \citep{Schmied2015} used a simplified prediction model with sinusoidal functions to predict the preceding vehicle behavior, and demonstrated a predictive cruise control approach. In \citep{Moser2015}, a Bayesian network approach with CLG model was used to predict the information of a target vehicle for the cooperative adaptive cruise control (CACC) use case. In \citep{Jia2020} long-short term memory (LSTM)-based energy-optimal ACC for target vehicle speed prediction in an urban environment is proposed. Furthermore, the authors in \citep{Wegener2021} implemented vector autoregressive (VAR) model to generate simultaneous predictions of the target vehicle, and used receding horizon control to derive optimal accelerations for the host vehicle.  

In most of the previous works \citep{Lin2018, Shin2019, Liu2019, Jia2020, Wegener2021}, the speed prediction models were trained on datasets that were gathered from repeated trials in the same driving route by considering either limited or no surrounding traffic. Although the driving patterns are comparatively easy to predict in such an approach, the limitation, however, remains that the prediction model may not generalize well for unseen data from a different route. To address this, the present work proposes a scalable and more generalizable method for preparing the time series data, and develop a speed predictor that uses the historical observations to predict the target vehicle future velocities in both urban and highway environments.  

The contributions made in this paper are:
(1) A novel approach for time series data preparation for urban and highway networks using the microscopic traffic simulation tool SUMO is presented. 
(2) To predict the target vehicle future velocities, both deep recurrent neural networks (LSTM and GRU) and physics-based models (CV and CA) are studied.  
(3) The influence of various input variables (e.g. preceding vehicle behavior, traffic light signal phase and road speed limits) on the prediction accuracy are investigated. Moreover, the impact of using additional V2V and V2I information on the accuracy of the predicted outputs is explored. 
(4) Furthermore, the performance of the EACC on the predicted target vehicle velocities is evaluated, and the energy-saving potential for different prediction horizons are investigated. 

\section{Ecological Adaptive Cruise Control}
\label{sec:EACC}
\subsection{System Dynamics}
\label{sec:System_Dynamics}
The longitudinal vehicle dynamics of the host car is described by
\begin{align}
\frac{dv_\text{h}}{dt} = \frac{1}{m_\text{eq}}(F_\text{t} - F_\text{b} - \underbrace{F_\text{a} - F_\text{roll} - F_\text{g}}_{F_\text{r}})
\label{eqn:Acceleration_1}
\end{align} 
where $v_\text{h}$ is the host car velocity, $m_\text{eq}$ is the equivalent mass which is the sum of vehicle weight, rotational equivalent masses, driver and cargo weight,  $F_\text{t}$ is the traction force, $F_\text{b}$ is the braking force and $F_\text{r}$ is the combination of aerodynamic resistance $F_\text{a} = \frac{1}{2}\rho A_\text{f} c_\text{a} v_\text{h}^2$, rolling resistance $F_\text{roll}=c_\text{r}m_\text{v}g \cos\theta$ and gradient resistance $F_\text{g} = m_\text{v}g \sin\theta$. To handle the nonlinearity occurring due to the term $v_\text{h}^2$ in $F_\text{a}$, an approximation of the aerodynamic resistance $F_\text{a} \approx \frac{1}{2}\rho A_\text{f} c_\text{a} (p_{1}v_\text{h}+p_\text{2})$ is considered in this work. Here, $c_\text{a}$ is the drag coefficient, $A_\text{f}$ is the frontal cross-sectional area of the vehicle, $\rho$ is the density of the air, $p_\text{1}$ and $p_\text{2}$  are the coefficients obtained through line fitting. Furthermore, $m_\text{v}$ is the host vehicle weight, $g$ is the gravitational acceleration, $\theta$ is the gradient angle and $c_\text{r}$ is the rolling resistance coefficient.

The chosen host vehicle in this work is a battery electric vehicle (BEV), whose accurate vehicle and battery models are obtained from \citep{Lin2014}. A half-map approximation of the BEV power consumption map \citep{Chada2020} is used in this study. 
\subsection{Model Predictive Control Problem Formulation}
\label{sec:problem_formulation}
\begin{figure}[t]
	\centering
	\includegraphics[width=0.8\linewidth]{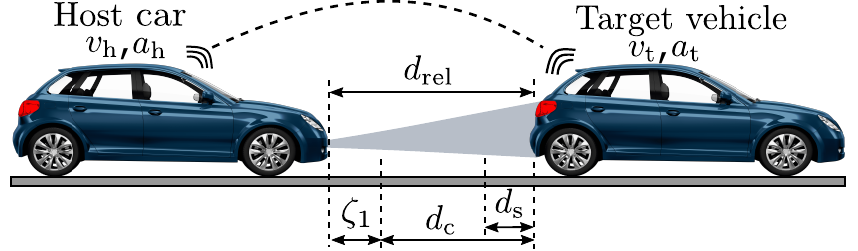}
	\caption{Schematic of a typical car-following scenario}
	\label{fig:Car-following scenario}
\end{figure}

A typical car-following scenario is illustrated in Fig.~\ref{fig:Car-following scenario}, in which a host car is tracking a target vehicle. The EACC optimization problem based on model predictive control (MPC) framework is formulated in time-domain with a goal to minimize the objective function (\ref{eq:MPC_Cost_Function}). Here, $N$ is the length of the prediction horizon and $k$ denotes the discrete time. 
\begin{subequations}
	\begin{flalign} 
	\label{eq:MPC_Cost_Function}
	& \min_{\textit{\textbf{$F_{t,k}$}},\textbf{$F_{b,k}$},{\zeta}_{1,k},{\zeta}_{2,k}}
	\resizebox{0.55\hsize}{!}{$\sum_{k=0}^{N-1} P(v_{\text{h},k}, F_{\text{t},k}) + \varepsilon_{1}F_{\text{b},k}^2 + \nonumber$} \\ & 
	\resizebox{0.85\hsize}{!}{$\ \ \ \ \ \ \ \ \ \ \ \ \ \ \ \ \ \ \ \ \ \ \ \ \ \ \ \ \ \ \ \ \ \ \ \ \ \ \ \ \ \ \ \ \ \varepsilon_{2}\zeta_{\text{1},k}^2 + \varepsilon_{3}\zeta_{\text{2},k}^2 $} \\ \label{eq:Online_Computation_4} 
	& \resizebox{0.9\hsize}{!}{$\text{s.t.} \ \  v_{\text{h},k+1}= v_{\text{h},k} + \frac{\Delta T }{m_\text{eq}}( F_{\text{t},k}- F_{\text{b},k}  - c_\text{r}m_\text{v}g\cos\theta_{k}$} - \nonumber \\ & 
	\resizebox{0.85\hsize}{!}{$\ \ \ \ \ \ \ \ \ \ \ \ \ \ \ \ \ \ \ \ \ \ \ \ \frac{1}{2}\rho A_\text{f}  c_\text{d} (p_\text{1} v_{\text{h},k}+ p_\text{2}) - m_\text{v}g\sin\theta_{k})$}  \\ \label{eq:Online_Computation_3}
	& 
	\resizebox{0.85\hsize}{!}{$d_{\text{rel},k+1}=d_{\text{rel},k}+ \Delta T \left( \frac{v_{\text{t},k}+v_{\text{t},k+1}}{2} -\frac{v_{\text{h},k}+v_{\text{h},k+1}}{2}\right)$}\\ \label{eq:Online_Computation_5}
	&\resizebox{0.6\hsize}{!}{$ \ \ \ \ \ \ \ \ \ \ \ \ \ \ \ v_{\text{min},k} \leq v_{\text{h},k} \leq  v_{\text{max},k}$} \\ \label{eq:Online_Computation_7} 
	& \resizebox{0.55\hsize}{!}{$ \ \ \ \ \ \ \ \ \ \ \ \ \ \ \ \ \ \ \ \ \ 0 \leq F_{\text{t},k} \leq  F_\text{t,max}$} \\	\label{eq:Online_Computation_8}   
	& \resizebox{0.55\hsize}{!}{$ \ \ \ \ \ \ \ \ \ \ \ \ \ \ \ \ \ \ \ \ \ \ 0 \leq F_{\text{b},k} \leq  F_\text{b,max}$} \\	\label{eq:Online_Computation_2} 
	&\resizebox{0.65\hsize}{!}{$  \ \ \ \ \ \ \ \ \ \ \ \ \ \ \ \  F_{\text{t},k}- F_{\text{t},k+1} - \zeta_{\text{2},k} \leq \Delta F_\text{t,max}$} \\	\label{eq:Online_Computation_9}
	& \resizebox{0.65\hsize}{!}{$ \ \ \ \ \ \ \ \ \ \ \ \ \ \ \ F_{\text{t},k+1} - F_{\text{t},k} - \zeta_{\text{2},k} \leq \Delta F_\text{t,max}$} \\	\label{eq:Online_Computation_10}	      
	& \ \ \ \ \ \ \ \ \ \ \ \ \ \ \ d_{\text{rel},k} \geq \underbrace{d_\text{min} + h_\text{m} v_{\text{h},k}}_{d_{\text{s},k}} \\ \label{eq:Online_Computation_6} 
	& \ \ \ \ \ \ \ \ \ \  d_{\text{rel},k} \leq \underbrace{ d_\text{min} + h_\text{c} v_{\text{h},k}}_{d_{\text{c},k}} + \zeta_{\text{1},k}
	\end{flalign}
	\label{eq:Online_Computation}  
\end{subequations}
 The first term in the cost function minimizes the BEV power consumption that can be approximated as a function of $v_{h,k}$ and $F_{t,k}$ \citep{Chada2020}. Using the second term in (\ref{eq:MPC_Cost_Function}), an excessive braking force $F_{\text{b},k}$ is penalized. In the third term, to motivate the host car to stay within the desired region $d_\text{c}$ to a target vehicle, a slack variable $\zeta_{1}$ is penalized. Furthermore, in the final term the variation in the traction force at successive time steps is penalized using a slack variable $\zeta_{2}$ if it exceeds a constant value $\Delta F_\text{t,max}$, with the aim to minimize jerks and improve the driving comfort, as given in (\ref{eq:Online_Computation_2} and \ref{eq:Online_Computation_9}). $\varepsilon_{1}$, $\varepsilon_{2}$ and $\varepsilon_{3}$ are the corresponding weighting factors for the aforementioned terms in the cost function (\ref{eq:MPC_Cost_Function}). The host vehicle velocity and relative distance to the target vehicle in discretized form is given in (\ref{eq:Online_Computation_4}) and  (\ref{eq:Online_Computation_3}), where $v_\text{t}$ is the velocity of the target vehicle. Furthermore, the limits for the traction force $F_\text{t}$ are set using (\ref{eq:Online_Computation_7}), in which $F_\text{t,max}$ is the maximum traction force. The physical limitations of the vehicle with respect to velocity and braking force are addressed in (\ref{eq:Online_Computation_5}) and (\ref{eq:Online_Computation_8}). In order to maintain a safe distance to the target vehicle, a hard constraint is introduced in (\ref{eq:Online_Computation_10}). Here, the state variables are $x_{k}= [v_{\text{h},k}, d_{\text{rel},k}]^\top$ and the control variables are $u^{*}_{k}= [F_{\text{t},k},F_{\text{b},k},\zeta_{\text{1},k},\zeta_{\text{2},k}]^\top$.
\section{Data Preparation for Speed Prediction}
\label{sec:data_collection}
\begin{figure}[t]
	\begin{minipage}{0.48\linewidth}
		\subfloat[Landstuhl highway]{\label{subfig:a}\includegraphics[width=\linewidth,keepaspectratio]{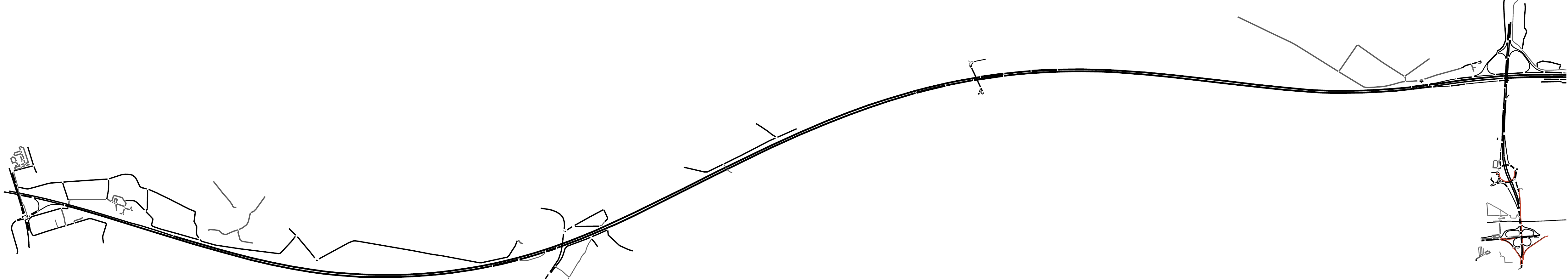}}\\
		\subfloat[Kaiserslautern city]{\label{subfig:b}\includegraphics[width=\linewidth,keepaspectratio]{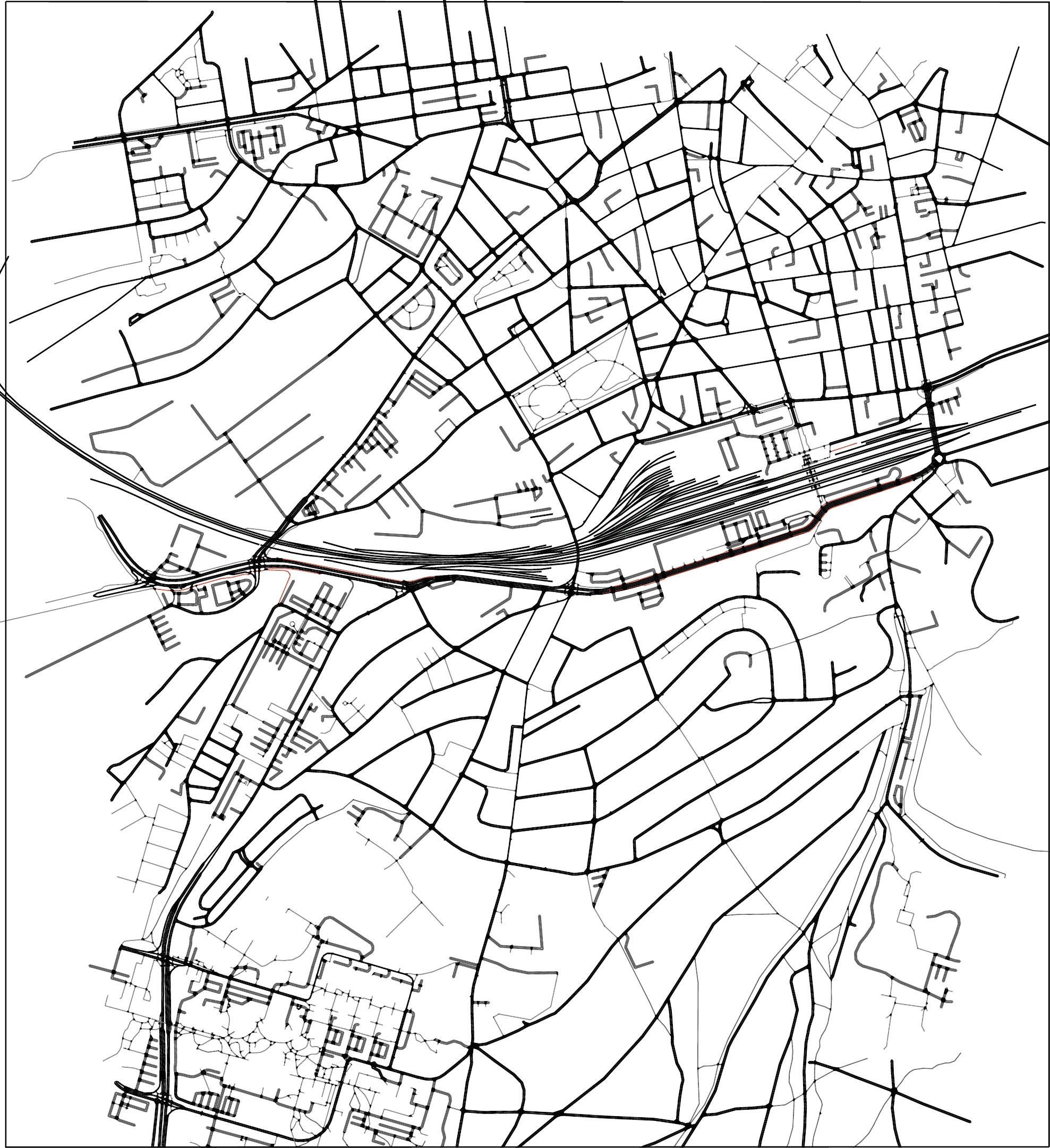}}
	\end{minipage}
	\hfill
	\begin{minipage}{0.48\linewidth}
		\subfloat[Nieder-Olm]{\label{subfig:c}\includegraphics[height=6.0cm,width=\linewidth]{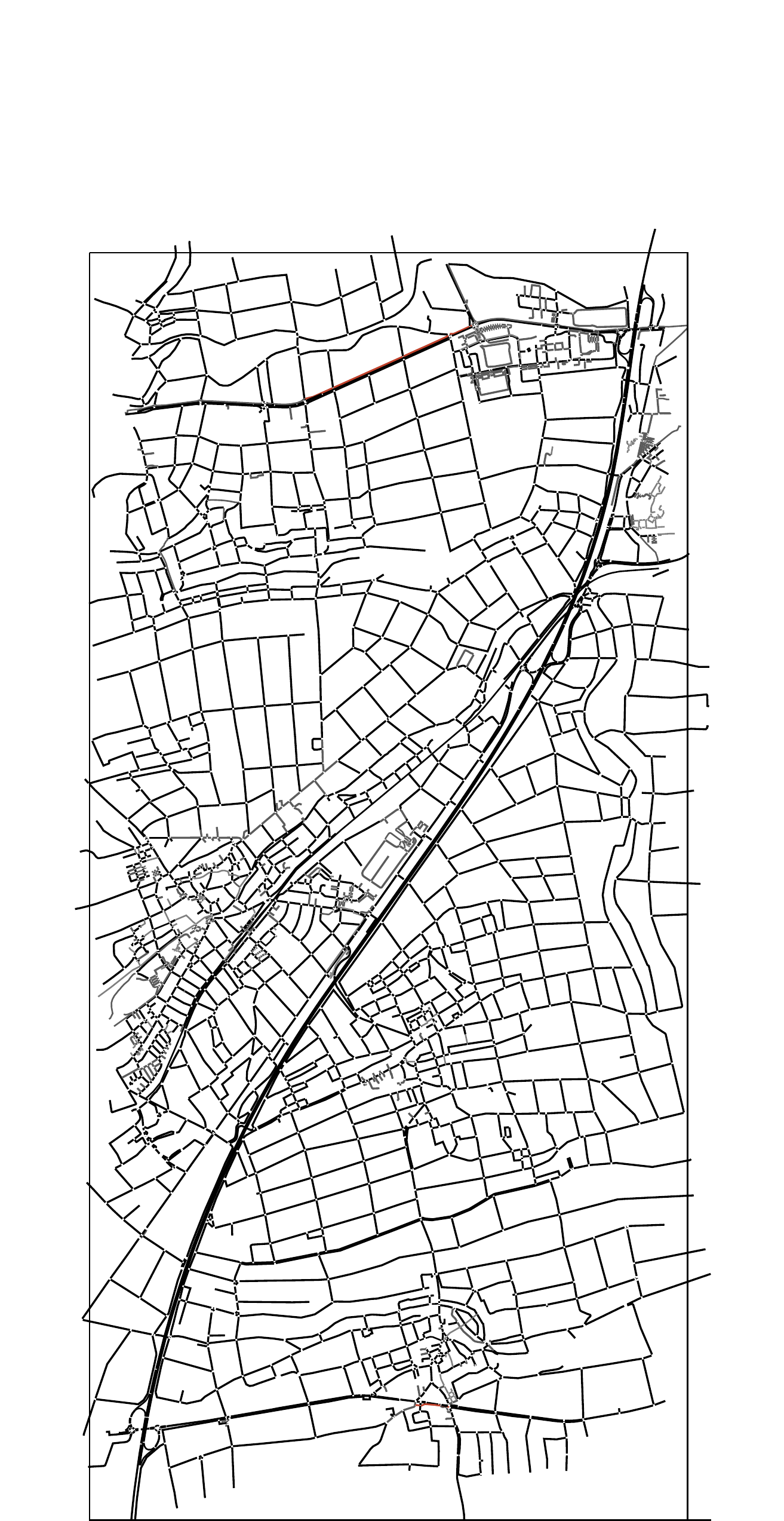}}
	\end{minipage}
	\caption{SUMO networks to generate training (a \& b) and testing (c) datasets}
\end{figure}

In contrary to the other approaches which considered a fixed driving route for data preparation, this work focuses on network-based data collection as it is laborious and time-consuming to extract V2V and V2I information in real world. The goal here is to gather rich driving datasets from the urban and highway networks that are route-independent under varied traffic conditions. Moreover, the data must enable developing scalable prediction models and promote generalizability. Therefore, to extract time series information in this work, the open source microscopic traffic simulation tool SUMO (Simulation for Urban Mobility)\citep{Lopez2018} was used. The road networks in SUMO are generated from the open-source map database \citep{OpenStreetMap}. To reproduce the real world traffic flow in the simulation, a random traffic ($3000$ vehicles/network) is generated with each traffic object in the network having a random starting and destination points \citep{Chada2021}. The training datasets are chosen from the Landstuhl highway (Fig.~\ref{subfig:a}) and Kaiserslautern city (Fig.~\ref{subfig:b}). To improve generalizability, five test datasets are chosen from a different city and connecting highway (Nieder-Olm), as shown in Fig.~\ref{subfig:c}.

In a highway driving scenario, the speed of the target vehicle is primarily influenced by the regulatory road speed limits and the velocities of a preceding vehicle. Besides these factors, the presence of traffic light signals in the urban environments influence the target vehicle speed as well. By leveraging the V2V and V2I information, additional inputs from multiple vehicles ahead of the target vehicle can be obtained. Moreover, the future signal phase and timing (SPaT) information can be used to improve the predictions for the target vehicle. 

\begin{figure}[t]
	\centering
	\includegraphics[width=1.0\linewidth]{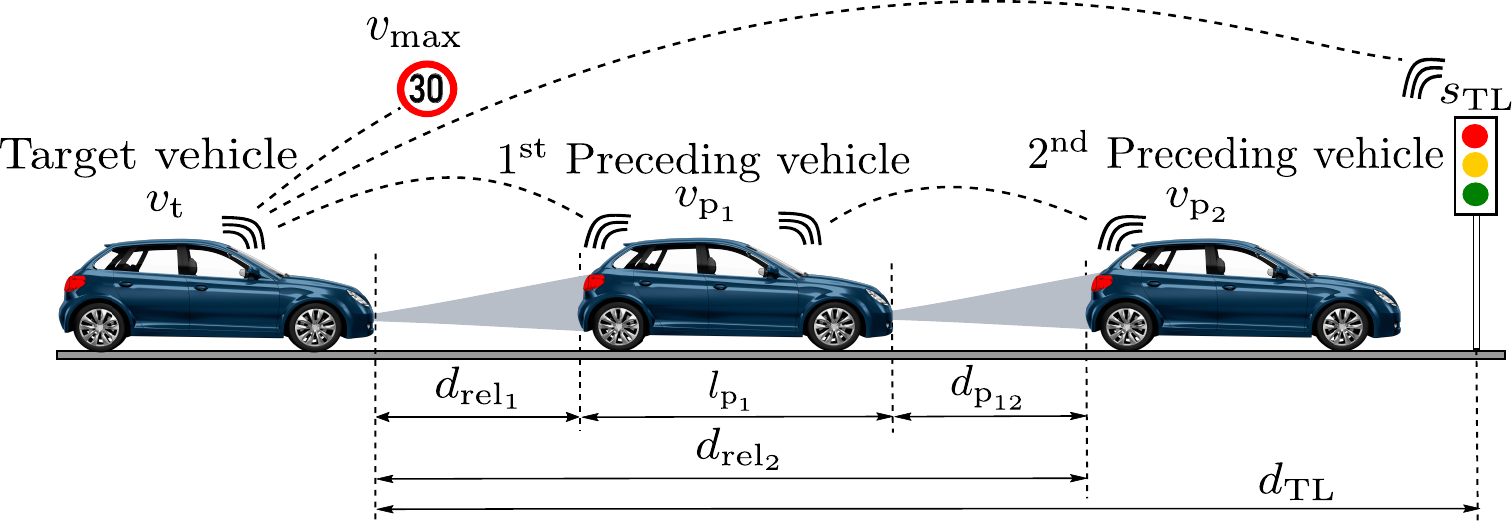}
	\caption{Schematic of V2V and V2I enabled traffic scenario}
	\label{fig:traffic_scenario}
\end{figure}

In this work, two feature groups FG1 and FG2 are investigated. As listed in Table~\ref{table:input_groups} and illustrated in Fig.~\ref{fig:traffic_scenario}, the feature group FG1 consists of six input features, namely, velocity of the target vehicle $v_{\text{t}}$, velocity of the first preceding vehicle $v_{\text{p}_1}$, relative distance between the target vehicle and the first preceding vehicle $d_{\text{rel}_1}$, traffic light signal current state $s_{\text{TL}}$, relative distance between the target vehicle and the traffic light signal $d_{\text{TL}}$, and the maximum road speed limit $v_{\text{max}}$. In addition to the input features described in FG1, the feature group FG2 considers the influence of the second preceding vehicle with velocity $v_{\text{p}_2}$, and the relative distance between the target vehicle and the second preceding vehicle $d_{\text{rel}_2}$ that is calculated using,
\begin{equation}
d_{\text{rel}_{2}} =\ d_{\text{rel}_{1}} +\ l_{\text{p}_{1}} +\ d_{\text{p}_{12}}
\end{equation}
where, $l_{\text{p}_{1}}$ is the length of the first preceding vehicle and $d_{\text{p}_{12}}$ is the relative distance between the first and second preceding vehicles. Besides, FG2 also considers the future traffic light signal state information $s_{\text{TL},k+1},..., s_{\text{TL},k+H}$ until the prediction horizon $H$ as input features. During the data collection process, each traffic object in the network is considered to be a target vehicle and the input variables as given in Table~\ref{table:input_groups} are gathered. To access the above mentioned variables from the simulation, a traffic control interface known as TraCI4Matlab is used \citep{Acosta2015}. Furthermore, preprocessing techniques such as data cleaning, normalization, and data splitting into training, validation and testing are performed to handle the data in an efficient manner.  
\section{Speed Prediction Methods}
\label{sec:methodology}
In this work, to develop a target vehicle speed predictor, both the physics-based prediction methods and deep recurrent neural networks are studied. 
\subsection{Physics-Based Prediction Models}
\label{sec:physics}
\subsubsection{Constant Velocity:}
\label{sec:CV}
The constant velocity (CV) model assumes that the future velocities of the target car remain constant, and can be determined using  
\begin{equation}
v_\text{t}( t+\Delta T)=\overline{v}_\text{t}(t)
\end{equation}
where $\overline{v}_\text{t}(t)=\frac{v_\text{t}(t)+v_\text{t}(t-1)}{2}$, $\Delta T$ is the sample time.
\subsubsection{Constant Acceleration:}
\label{sec:CA}
The constant acceleration (CA) model makes the assumption that the future velocities of the target car are incremented by a constant amount of acceleration, i.e. 
\begin{equation}
v(t+\Delta T)=v(t)+\Delta Ta(t)
\end{equation}
where $a(t)=\frac{v_\text{t}(t)-v_\text{t}(t-1)}{\Delta T}$. 
\begin{table}[t]
	\centering
	\caption{Feature groups and their corresponding input variables}\label{table:input_groups}
	\begin{tabular}{cc} 
		\hline
		\textbf{Feature group}        & \textbf{Input Features}                                                                                      \\ 
		\hline\hline
		FG1                  & $v_{\text{t},k}, v_{\text{p}_1,k}, d_{\text{rel}_1,k}, s_{\text{TL},k}, d_{\text{TL},k},v_{\text{max},k}$               \\ 
		\hline
		\multirow{2}{*}{FG2} & $v_{\text{t},k}, v_{\text{p}_1,k}, d_{\text{rel}_1,k}, s_{\text{TL},k}, d_{\text{TL},k},v_{\text{max},k}$              \\
		& \multicolumn{1}{l}{$v_{\text{p}_2,k}, d_{\text{rel}_2,k}, s_{\text{TL},k+1},..., s_{\text{TL},k+H}$}  \\
		\hline\hline
	\end{tabular}
\end{table}
\subsection{Recurrent Neural Networks}
\begin{figure}[t]
	\centering
	\includegraphics[width=0.7\linewidth]{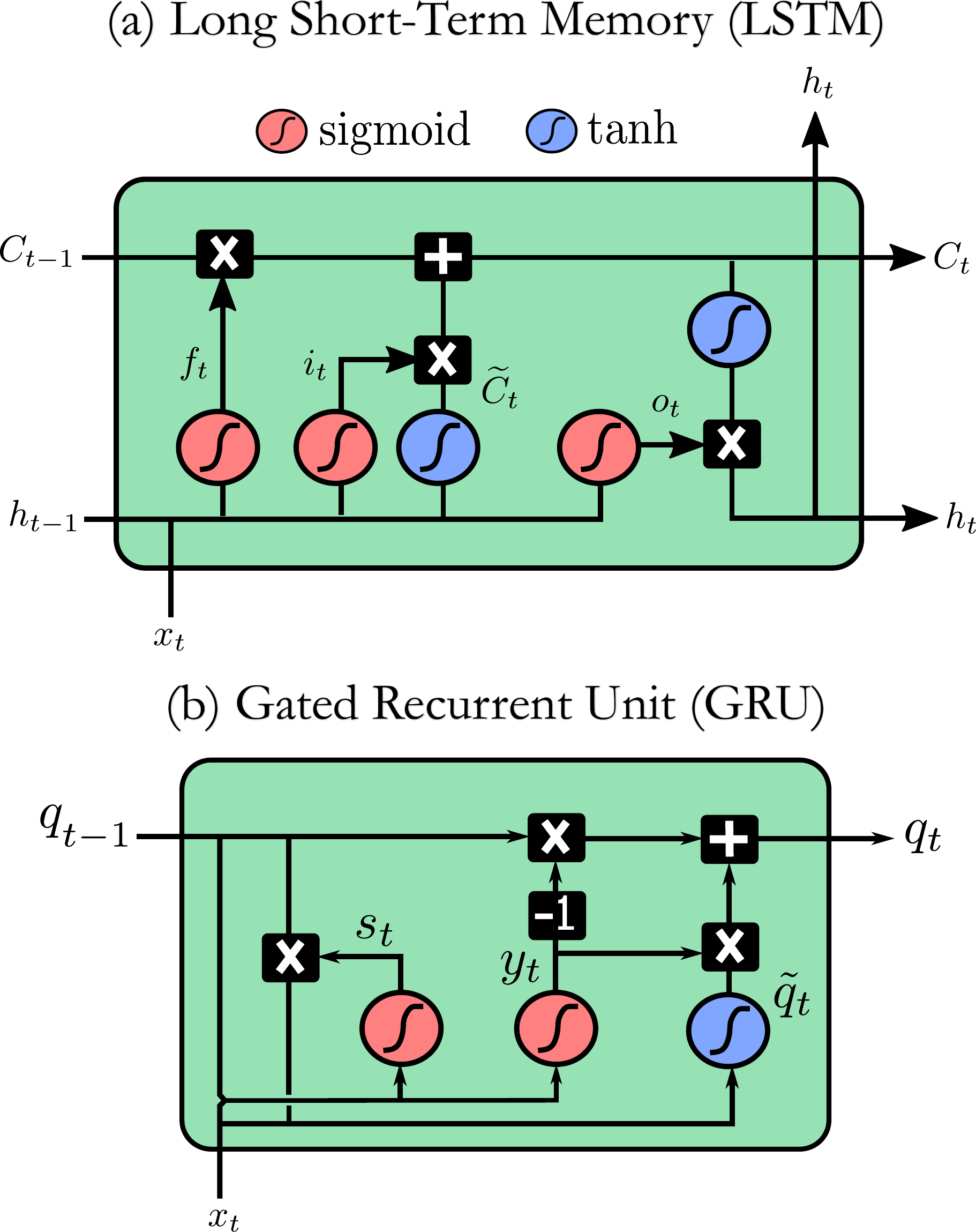}
	\caption{Internal cell structure of LSTM and GRU}
	\label{fig:lstm_architecture}
\end{figure}

Recurrent neural networks (RNNs) are popularly known in the category of deep neural networks as they are capable of using their internal state memory to process sequential or time series data. Two variants in the RNN architecture, namely, gated recurrent unit (GRU) and long short-term memory (LSTM) units are investigated in this work. The internal cell structure of LSTM and GRU are illustrated in the Fig.~\ref{fig:lstm_architecture}(a) and Fig.~\ref{fig:lstm_architecture}(b), respectively. 

The internal mechanisms of an LSTM (Fig.~\ref{fig:lstm_architecture}(a)) consists of a cell state $C_\text{t}$ and various gates such as a forget gate $f_\text{t}$, an input gate ($i_\text{t}$, $\!\widetilde C_\text{t}$) and an output gate $o_\text{t}$. The cell states act as a transport highway to transfer the information from previous intervals all the way down the entire sequence chain. The gates can regulate the flow of information by either adding or removing the information from the cell state. The inputs to the LSTM are the previous hidden state $h_\text{t-1}$ and the current state $x_\text{t}$. The forget gate uses a sigmoid activation function to decide on which information to throw away from the cell state, and is described using (\ref{eq:LSTM_equations_1}).  
\begin{subequations}
	\begin{flalign} 
	\label{eq:LSTM_equations_1}
	& f_\text{t} = \sigma \left( W_{f}\left[ h^{T}_{t-1} ,X^{T}_{t}\right] +b_{f}\right) \\ \label{eq:LSTM_equations_2}
	&i_\text{t} = \sigma \left( W_{i}\left[ h^{T}_{t-1} ,X^{T}_{t}\right] +b_{i}\right)\\ \label{eq:LSTM_equations_3}
	& \!\widetilde C_\text{t} = \text{tanh}\left( W_{C}\left[ h^{T}_{t-1} ,X^{T}_{t}\right]^{T} +b_{C}\right)\\ \label{eq:LSTM_equations_4}
	& C_\text{t} = f_{t} *C_{t-1} +i_{t} *\!\widetilde C_\text{t}\\ \label{eq:LSTM_equations_5}
	&o_\text{t} = \sigma \left( W_{o}\left[ h^{T}_{t-1} ,X^{T}_{t}\right] +b_{o}\right)\\ \label{eq:LSTM_equations_6}
	& h_\text{t} = o_{t} *\text{tanh}( C_{t})
	\end{flalign}
	\label{eq:LSTM_equations}  
\end{subequations}
The input gate is a combination of two layers. The first layer is referred to as input layer gate $i_\text{t}$, which decides on which important information to retain to update in the cell-state and the second layer is a tanh layer which adds new candidate values $\!\widetilde C_\text{t}$ to the cell state as given in (\ref{eq:LSTM_equations_2}) and (\ref{eq:LSTM_equations_3}) respectively. Thus, the old cell state $C_\text{t-1}$ is updated to a new cell state $C_\text{t}$ using the forget gate and the input gate information according to (\ref{eq:LSTM_equations_4}). Finally, the cell state passes through the tanh activation function and the final output is filtered using a sigmoid layer resulting in the next hidden state $h_\text{t}$ as described in (\ref{eq:LSTM_equations_5}) and (\ref{eq:LSTM_equations_6}). In equation (\ref{eq:LSTM_equations}), $W_\text{f}$, $W_\text{i}$,$W_\text{c}$, $W_\text{o}$ are the weights and $b_\text{f}$, $b_\text{i}$,$b_\text{c}$, $b_\text{o}$ are the biases of the forget, input and output gates respectively. 

In contrary to the LSTM, the GRU has a simpler cell structure with only two gates, namely, reset and update gates, as illustrated in Fig.~\ref{fig:lstm_architecture}(b). The reset gate is used to determine how much historical information to forget and the update gate decides which new information must be passed along to the future. The equations for the GRU are given by
\begin{subequations}
	\begin{flalign} 
	\label{eqn:2.21}
	& y_t=\sigma(U_yx_t+W_yq_{t-1}+a_y) \\ \label{eqn:2.22} 
	& s_t=\sigma(U_sx_t+W_sq_{t-1}+a_s) \\ \label{eqn:2.23} 
	& \tilde q_t=tanh(U_q x_t+W_q(s_t \odot q_{t-1}+a_q)) \\ \label{eqn:2.24} 
	& q_t=(1-y_t)\odot q_{t-1}+y_t \odot \tilde{q}_t
	\end{flalign}
	\label{eq:d_safe_comfort}  
\end{subequations}
where $U_{*}$, $W_{*}$, and $a_{*}$ are the two weights and the bias, while $\odot$ is the scalar product of the vectors and $\sigma$ refers to the sigmoid activation function.

\begin{figure}[t]
	\centering
	\includegraphics[width=0.9\linewidth]{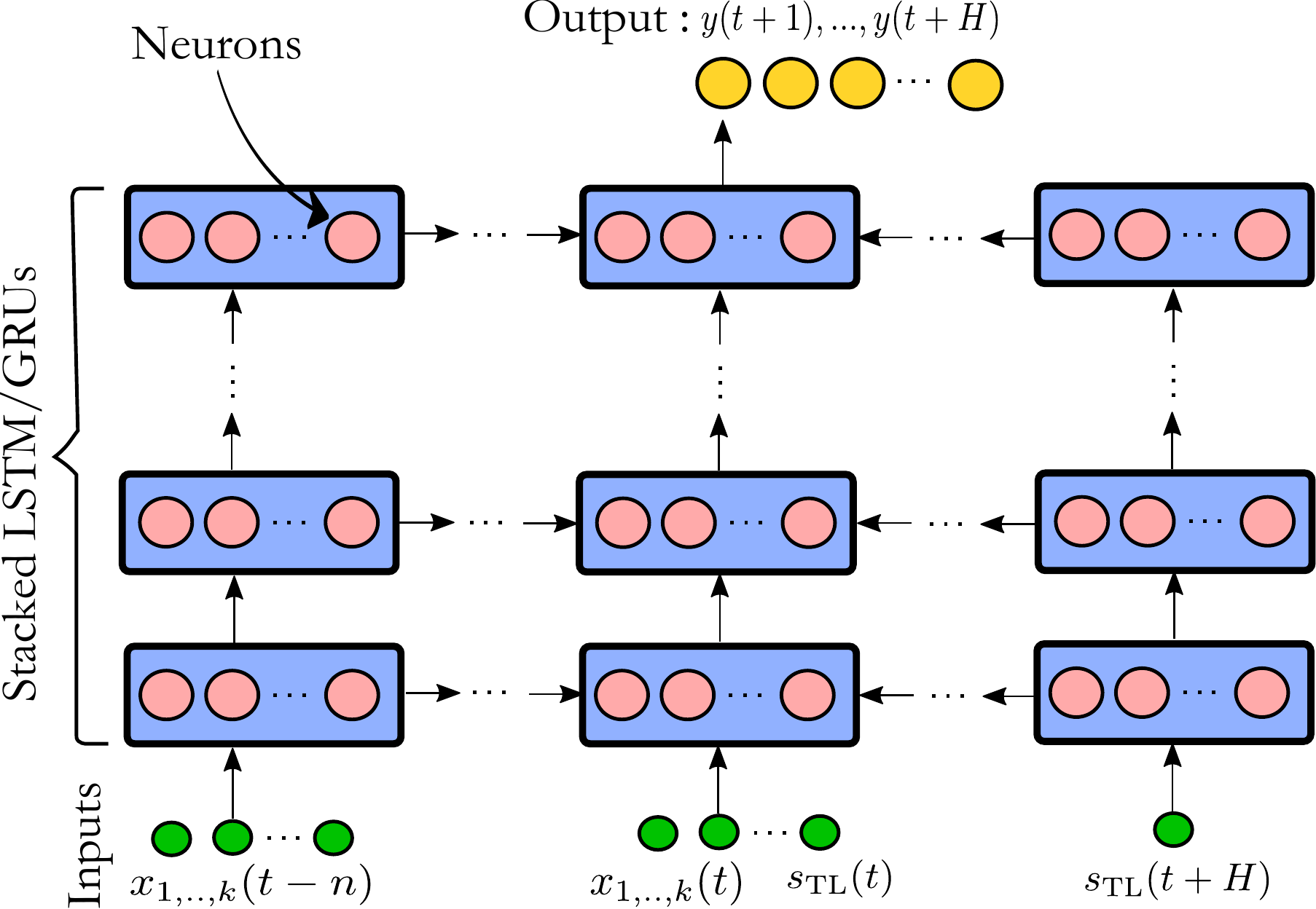}
	\caption{Architecture of Stacked RNN}
	\label{fig:lstm}
\end{figure}
\begin{figure*}[t]
	\centering
	\includegraphics[width=0.8\linewidth]{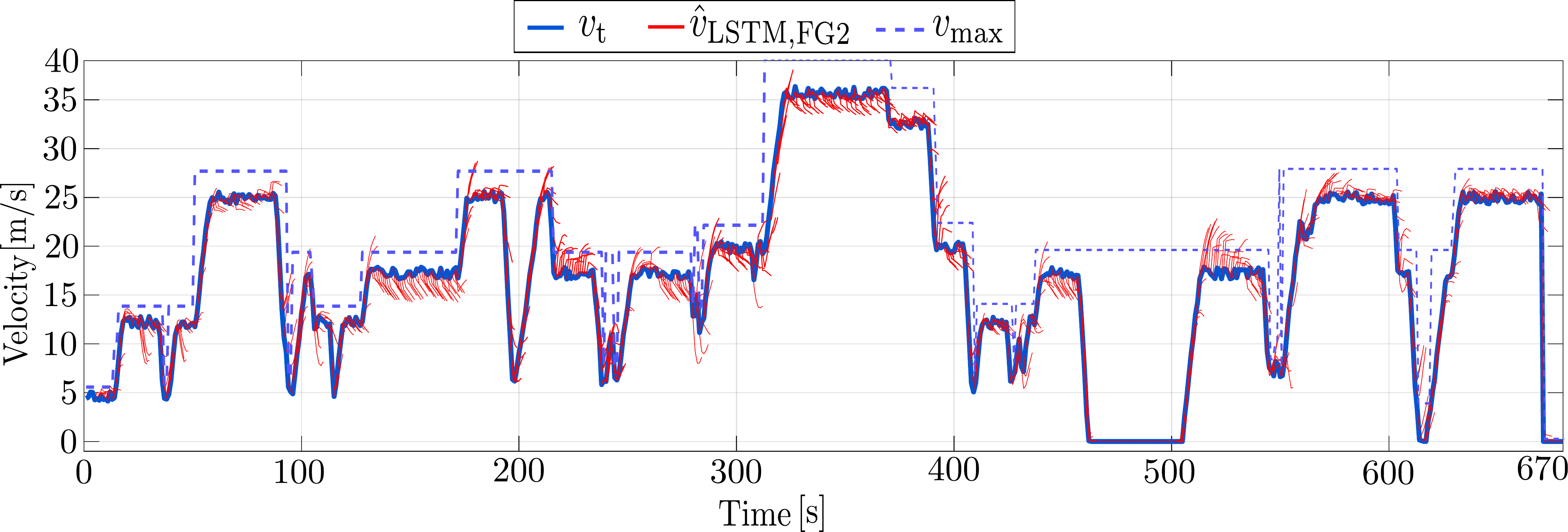}
	\caption{Prediction results of LSTM-FG2 on one of the test datasets for a prediction horizon of $5\,\mathrm{s}$}
	\label{fig:IG2_5s}
\end{figure*}
\begin{figure*}[t]
	\centering
	\includegraphics[width=0.8\linewidth]{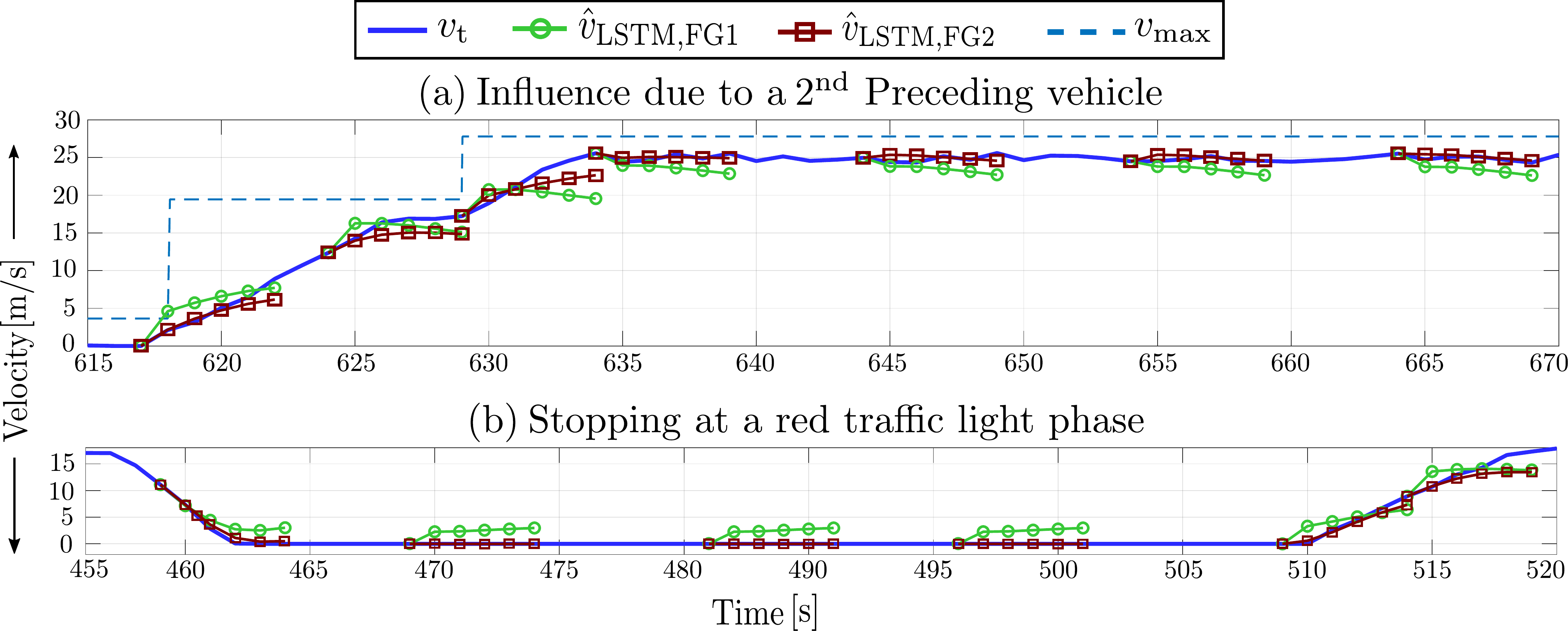}
	\caption{Comparison of target vehicle predictions for LSTM-FG1 and FG2 models for a prediction horizon of $5\,\mathrm{s}$ }
	\label{fig:IG1_vs_IG2_5s}
\end{figure*}
\subsubsection{Stacked RNN:}
In order to obtain a greater level of abstraction, the hidden layers are arranged together in stacked fashion as illustrated in Fig. \ref{fig:lstm}. The architecture consists of three segments, described from bottom to top: (i) an input layer, (ii) stacked LSTM or GRU layers, (iii) an output layer. The input layer consists of multiple input features (marked in green) as described in Table~\ref{table:input_groups}. In Fig. \ref{fig:lstm} from left to right, the feature values of past, current and future timesteps are provided as inputs to the stacked layers. On the right, along with the current traffic light state, future state information up to a prediction horizon $H$ is additionally used. The combined input sequence is fed into the stacked hidden layers. Each layer has memory cells associated with it, otherwise known as neurons. The lower layers are used to capture the low level representations and the deeper layers can learn higher levels of abstraction. The layer in yellow outputs multi-step sequence predictions up to a prediction horizon $H$, in this case the future velocities of the target vehicle.  

To model the RNNs in this work, the keras deep learning library was used \citep{Jason2018} in python, and training of the deep learning prediction models is performed on a GPU cluster. The hyperparameters for both the LSTM and GRU are tuned using the Keras Bayesian optimization function. The tuned hyperparameters for the $5\,\mathrm{s}$ speed predictors are illustrated in the Table \ref{table:hyper_LSTM_GRU}. Four stacked layers were implemented to model the non-linear input data representations to outputs. Moreover, to reduce overfitting of the data during the training process, a regularization technique called dropout is used. Furthermore, a rectified linear activation function (reLU) is used to train the network faster, and to estimate the loss of the model while training, a mean squared error (MSE) is used as the loss function. Additionally, the ADAM optimizer is used to efficiently update the network weights while training. The length of the past sequence is maintained to be equivalent to the prediction horizon $H$, and the learning rate is chosen as $1e^-3$. The number of epochs was chosen as 25, in addition to which an early stopping was implemented to be able to stop the training process when a desired metric has stopped to improve.
\begin{table}[]
	\begin{tabular}{c|cccc}
		\hline
		\multirow{2}{*}{\textbf{Hyperparameters}} & \multicolumn{2}{c|}{FG1}                 & \multicolumn{2}{c}{FG2} \\ \cline{2-5} 
		& LSTM & \multicolumn{1}{c|}{GRU} & LSTM   & GRU  \\ \hline \hline
			Batch size                                & \multicolumn{2}{c}{32}                            & \multicolumn{2}{c}{512}          \\
		Input features                            & \multicolumn{2}{c}{6}                             & \multicolumn{2}{c}{13}           \\ 
		Stacked layer 1                           & 90            & 450                               & 600             & 300            \\
		Dropout 1                                 & 0             & 0.3                               & 0               & 0.2            \\
		Stacked layer 2                           & 60            & 600                               & 420             & 600            \\
		Dropout 2                                 & 0             & 0.3                               & 0.25            & 0              \\
		Stacked layer 3                           & 600           & 60                                & 450             & 600            \\
		Dropout 3                                 & 0.3           & 0                                 & 0.3             & 0              \\
		Stacked layer 4                           & 600           & 60                                & 480             & 180            \\
		Dropout 4                                 & 0             & 0.3                               & 0.3             & 0.1            \\
		Dense layer                               & 30            & 570                               & 60              & 30             \\
	\hline \hline
	\end{tabular}
	\caption{Hyperparameters for LSTM and GRU models for the predicition horizon of  $5\,\mathrm{s}$}
	\label{table:hyper_LSTM_GRU}
\end{table}
\section{Results and discussion}
\label{sec:results}
\begin{figure*}[t]
	\centering
	\includegraphics[width=0.7\linewidth]{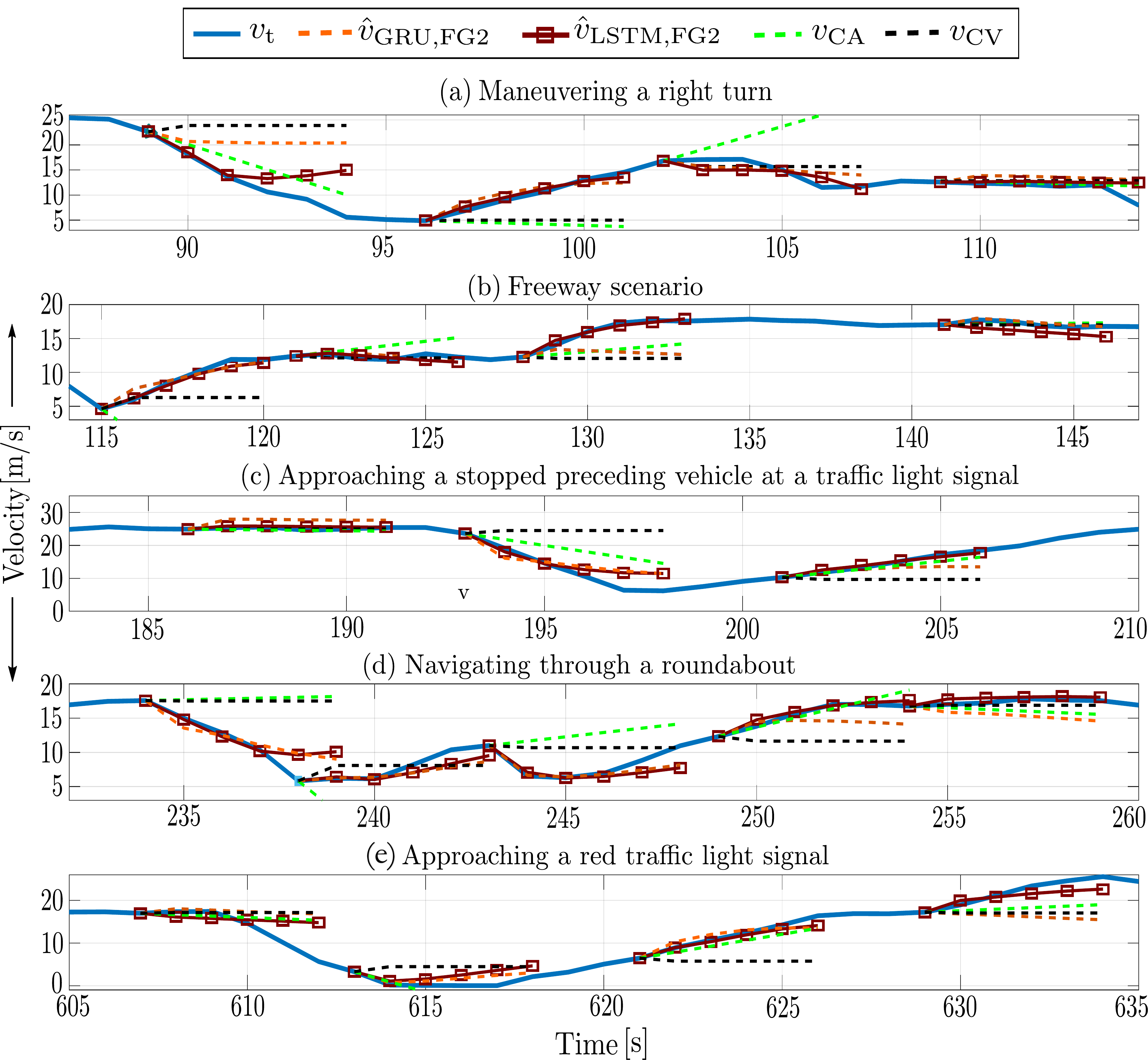}
	\caption{Speed forecasting results of $\hat{v}_{\text{GRU}\text{,FG2}}$, $\hat{v}_{\text{LSTM}\text{,FG2}}$, $\hat{v}_{\text{CA}}$ and $\hat{v}_{\text{CV}}$ for a prediction horizon of $5\,\mathrm{s}$}
	\label{fig:prediction_5f}
\end{figure*}
\subsection{Evaluation Metrics}
\label{sec:Evaluation_metric}
To evaluate the performance of the prediction models on the test dataset, two error metrics are used in this work. Firstly, the mean absolute error (MAE) measures the average magnitude of the errors in a set of predictions and is given by
\begin{equation}
\text{MAE}=\sum\limits ^{n}_{i=1}\frac{|\hat{v}_{t(i)} -v_{t(i)} |}{n}
\end{equation}
where $\hat{v}_{t(i)}$ is the predicted target vehicle speed at the $i^{th}$ prediction step, $v_{t(i)}$ is the original target vehicle observation and $n$ is the total number of observations. Secondly, the root mean squared error (RMSE) is the square root of the average of squared differences between predictions and actual observation and is described as
\begin{equation}
\text{RMSE}=\sqrt{\sum\limits ^{n}_{i=1}\frac{(\hat{v}_{t(i)} -v_{t(i)})^{2}}{n}}
\end{equation}
\subsection{Performance Evaluation of the Prediction Methods}
\label{sec:results_speed_prediction}
\begin{table}[h]
	\centering
	\resizebox{\columnwidth}{!}{\begin{tabular}{cllllllll} 
		\hline\hline
		\multicolumn{1}{c|}{\multirow{3}{*}{\begin{tabular}[c]{@{}c@{}}\textbf{Prediction}\\\textbf{Models }\end{tabular}}} & \multicolumn{4}{c|}{\textbf{MAE~[m/s]}}  & \multicolumn{4}{c}{\textbf{RMSE~[m/s] }}                                                                                                                            \\ 
		\cline{2-9}
		\multicolumn{1}{c|}{}                                                                                               & \multicolumn{2}{c|}{\textbf{5s }}                  & \multicolumn{2}{c|}{\textbf{ 10s}} & \multicolumn{2}{c|}{\textbf{5s }}                  & \multicolumn{2}{c}{\textbf{ 10s}}               \\ 
		\cline{2-9}
		\multicolumn{1}{c|}{}                                                                                               & \multicolumn{1}{c}{FG1} & \multicolumn{1}{c|}{FG2} & \multicolumn{1}{c}{FG1} & \multicolumn{1}{c|}{FG2}   & \multicolumn{1}{c}{FG1} & \multicolumn{1}{c|}{FG2} & \multicolumn{1}{c}{FG1} & \multicolumn{1}{c}{FG2} \\ 
		\hline\hline
		
		CV       & \multicolumn{2}{c}{2.12}                 & \multicolumn{2}{c}{3.16}                 & \multicolumn{2}{c}{3.79}                 & \multicolumn{2}{c}{5.43}\\
		CA       & \multicolumn{2}{c}{2.75}                 & \multicolumn{2}{c}{5.26}    & \multicolumn{2}{c}{4.77}                 & \multicolumn{2}{c}{8.68}               \\
		LSTM                                                                                                                 & 2.06                    & \textbf{1.75}            & 4.26                    & \textbf{2.92}   & 3.19                    & \textbf{2.84}            & 5.63                    & \textbf{4.61}    \\
		GRU                                                                                                                & 2.02                    & 2.51                    & 3.19                    & 3.3   & 3.29                    & 3.68                     & 4.58                    & 4.7                                \\
		\hline\hline
	\end{tabular}}
	\caption{Comparison of MAE and RMSE for various prediction methods}
	\label{tab:RMSE_comparison}
\end{table}
The performance of the prediction methods discussed in Section~\ref{sec:methodology} is evaluated on the $5$ test datasets. The average MAE and RMSE for each prediction model across different prediction horizons ($5\,\mathrm{s}$ and $10\,\mathrm{s}$) with respect to the feature groups (FG1 and FG2) are summarized in Table~\ref{tab:RMSE_comparison}. It can be noticed that the prediction error increases for all the models as the prediction horizon increases due to the increasing uncertainties. Moreover, the LSTM model with input features FG2 has outperformed remaining methods, and showcased lower average prediction error as compared to feature group FG1.  

The prediction results for the LSTM-FG2 model with a prediction horizon of $5\,\mathrm{s}$ after evaluating it on one of the test datasets that consists of various traffic scenarios are shown in Fig.~\ref{fig:IG2_5s}. Furthermore, the prediction result of $\hat{v}_\text{LSTM-FG1}$ is compared against $\hat{v}_\text{LSTM-FG2}$ in Fig.~\ref{fig:IG1_vs_IG2_5s} and demonstrated for two scenarios. In Fig.~\ref{fig:IG1_vs_IG2_5s}(a), the target vehicle is under the influence of two preceding vehicles ahead (exemplary scenario in Fig.~\ref{fig:traffic_scenario}). As presented in Table~\ref{table:input_groups}, the FG1 uses the inputs from the first preceding vehicle alone and the FG2 uses both the preceding vehicle information as inputs with an assumption that this information can be obtained from V2V communication. The results illustrate that the $\hat{v}_\text{LSTM-FG2}$ has better tracking ability of the target vehicle ${v}_\text{t}$ as compared to $\hat{v}_\text{LSTM-FG1}$. In the second scenario as illustrated in Fig.~\ref{fig:IG1_vs_IG2_5s}(b), the target vehicle stopped at a red phase of a traffic light signal. The predictions of $\hat{v}_\text{LSTM-FG2}$ match the expected behavior of the target vehicle ${v}_\text{t}$, however, as the feature group FG1 did not take the future traffic light phase into account, the $\hat{v}_\text{LSTM-FG1}$ depict inaccurate predictions that the target vehicle will move forward from standstill. Such a behavior may in turn lead to traffic signal violations which is not desirable.

Furthermore, a comparison of the prediction results for the methods discussed in Table~\ref{tab:RMSE_comparison} when evaluated on a few scenarios can be found in Fig.~\ref{fig:prediction_5f}. It can be observed that the LSTM-FG2 prediction model has demonstrated better prediction accuracy in all the scenarios as compared to GRU-FG2, CV and CA models. Although the prediction results of $\hat{v}_\text{GRU,FG2}$ matches with the predictions of $\hat{v}_\text{LSTM,FG2}$ at a few timestamps, its accuracy needs to be improved in a few scenarios (e.g. round-abouts). The predictions based on CV and CA models were not able to accurately predict the target vehicle behavior due to the presence of abrupt speed variations in the target vehicle. It can be noticed from Fig.~\ref{fig:prediction_5f}, that a prediction error exists with the $\hat{v}_\text{LSTM,FG2}$ as well. For instance, while maneuvering a right turn (Fig.~\ref{fig:prediction_5f}(a)), at $89\,\mathrm{s}$ the model did not predict the slow down at the curvature accurately. Similarly, in Fig.~\ref{fig:prediction_5f}(d) the model is able to predict the future speeds accurately in the round-about only until $3\,\mathrm{s}$ into the future.  
\subsection{Performance Evaluation of EACC using Predicted Speeds}
\label{sec:results_EACC}
\label{sec:conclusion}
\begin{figure*}[t]
	\centering
	\includegraphics[width=0.8\linewidth]{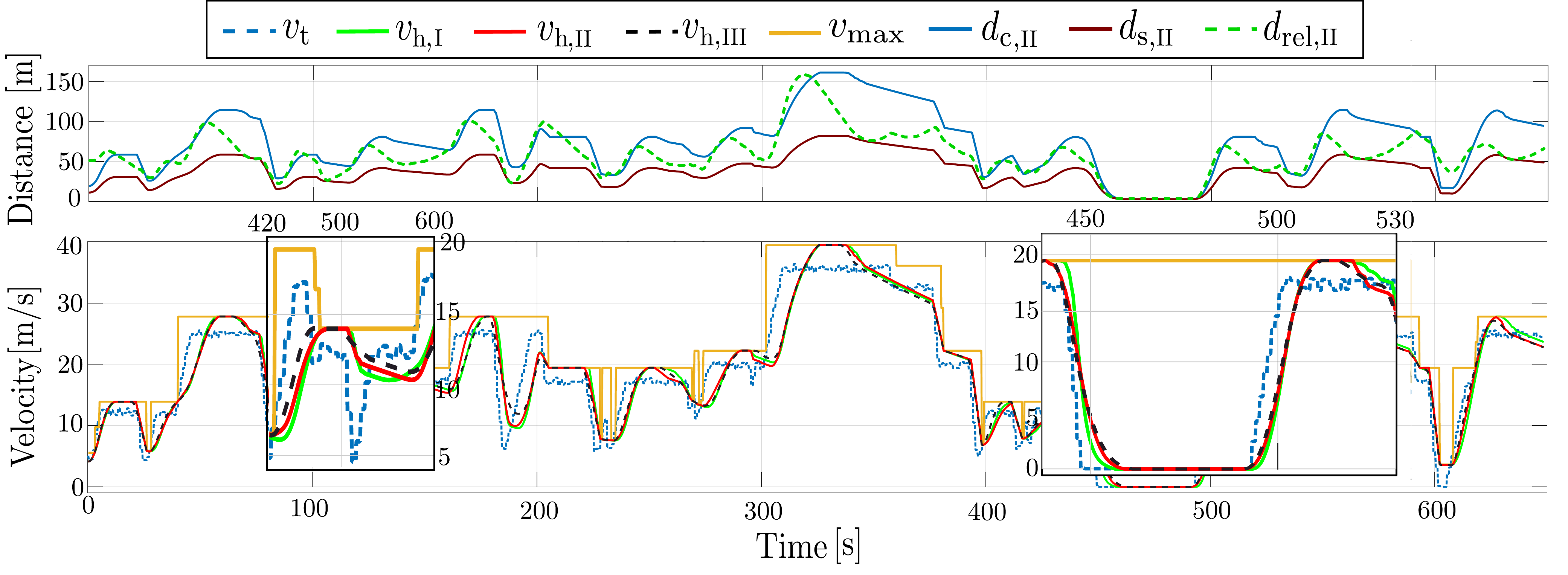}
	\caption{Speed profiles of EACC-equipped host car while tracking three target vehicle velocity criteria}
	\label{fig:MPC_result}
\end{figure*}
\begin{figure}[t]
	\centering
	\includegraphics[width=1.0\linewidth]{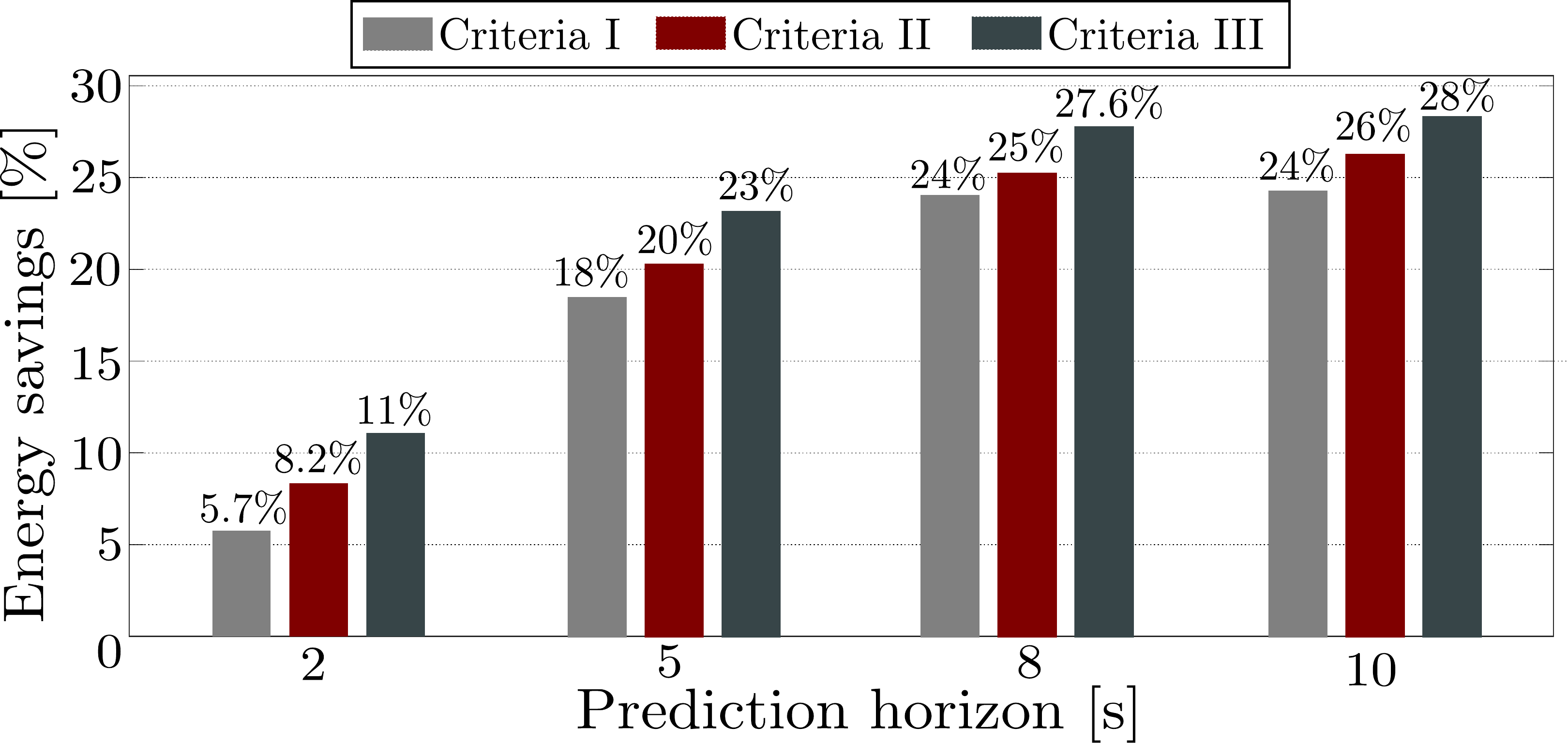}
	\caption{Energy savings for EACC-equipped host car while tracking three target vehicle velocity criteria}
	\label{fig:energy_savings_118}
\end{figure}
To analyze the performance of the EACC in a typical car-following scenario while tracking a target vehicle, an evaluation is conducted based on three criteria in which the target vehicle future velocities are (I) assumed to be constant due to the lack of a robust speed predictor and V2V communications, (II) predicted using the LSTM-based speed prediction model proposed in this work, and (III) assumed to be perfectly available. The speed profiles of the host vehicle for the three criteria $v_\text{h,I}$, $v_\text{h,II}$ and $v_\text{h,III}$ are illustrated in Fig.~\ref{fig:MPC_result} while tracking a target vehicle velocity profile $v_\text{t}$. In the distance plot in the above part of Fig.~\ref{fig:MPC_result}, the host vehicle with condition II is seen performing a robust car-following and maintains a good inter-vehicle distance to the target vehicle $d_\text{rel,II}$. It can be noticed from the two enlarged sections in the below part of Fig.~\ref{fig:MPC_result}, that the $v_\text{h,II}$ is able to track the $v_\text{h,III}$ better than the $v_\text{h,I}$. Moreover, considering a constant speed (CS) model in criterion I has resulted in abrupt decelerations and sharp accelerations as compared to the II and III.  

In Fig.~\ref{fig:energy_savings_118}, a comparison of the energy savings of the EACC-equipped host vehicle after evaluating on the three above mentioned criteria is shown. The results demonstrate that the energy savings can increase with the increase in the prediction horizon, i.e. the more the information about the target vehicle is available into the future, the better that the host vehicle is able to plan its optimal trajectories. Moreover, the proposed LSTM-based EACC (criterion II) is able to achieve better energy savings as compared to the criterion I and is close to criterion III.  

Concerning the execution time of the proposed LSTM-based EACC, an evaluation is performed for the prediction horizon of $10\,\mathrm{s} ~ (50\,\mathrm{steps})$ using Matlab$\textsuperscript{\textregistered}$ R2021b profiler on a Windows 10 PC equipped with an Intel$\textsuperscript{\textregistered}$ Core$\textsuperscript{\texttrademark}$ i7-7500U CPU processor with 2.70 GHz clock frequency and 12 GB RAM. No GPU for parallel computing was used. The mean execution time for the proposed concept is found to be $60\,\mathrm{ms}$. The sample time in this work is chosen as $\Delta T$=$200\,\mathrm{ms}$ including the time for inference of the neural networks. As with the current system configuration, the LSTM-based EACC can be solved at each step below the sample time, thus showcasing the real-time capability of the proposed controller. 

\section{Conclusion}

In this work, to enhance the efficiency of an ecological adaptive cruise control (EACC) strategy, an LSTM-based target vehicle speed prediction model for both urban and highway scenarios is proposed. In the speed prediction task, the LSTM model outperformed GRU, CV and CA models, and was able to capture the historical dependencies from several input features and perform long-term predictions up to $10\,\mathrm{s}$. Moreover, considering the additional input features, such as information about multiple preceding vehicles in the driving route obtained through V2V and traffic light signal future phases gathered through V2I has enhanced the prediction accuracy. Furthermore, energy savings up to $26\%$ can be realizable for the EACC-equipped host car while tracking the target vehicle predicted velocities. A performance increment in terms of additional average energy savings of up to $2.5\%$ with the proposed LSTM-based EACC can be achieved as compared to the constant speed (CS) model. For further improvements in the prediction accuracy, additional input features constituting traffic rules at non-priority intersections, road topology and curvature must be further investigated.   


\bibliography{ifacconf}             
                                                   







\end{document}